# Discrepancy-Sensitive Dynamic Fractional Cascading, Dominated Maxima Searching, and 2-d Nearest Neighbors in Any Minkowski Metric

Mikhail J. Atallah[*]   Marina Blanton[*]   Michael T. Goodrich[†]

Stanislas Polu[‡]

**Abstract**

This paper studies a discrepancy-sensitive approach to dynamic fractional cascading. We show, for example, that a search for a value $x$ in a collection of catalogs, of size at most $n$, stored in vertices of a path $P$ can be done in time $O(\log n + \sum_{(v,w) \in P} \log \delta_{v,w}(x))$, where $\delta_{v,w}(x)$ is the relative local discrepancy at $x$ of the catalogs stored at the nodes $v$ and $w$ in $G$. Such an approach is useful in real-world scenarios, for it leads to faster query and update times in many cases. We provide an efficient data structure for dominated maxima searching in a dynamic set of points in the plane, which in turn leads to an efficient dynamic data structure that can answer queries for nearest neighbors using any Minkowski metric. Specific bounds are derived for uniformly distributed data, and we also provide experimental results that show this discrepancy-sensitive approach works well in practice.

## 1 Introduction

*Discrepancy theory* deals with the degrees to which point sets differ from their expected uniformity (e.g., see Chazelle [9, 10]). This theory is usually applied globally, for entire sets, but we are interested in local notions of discrepancy, dealing with how sets differ from their expected uniformity in small intervals. This interest is motivated from *dynamic fractional cascading* [11, 12, 18].

---

[*]Dept. of Computer Sciences, Purdue Univ., {mja,mbykova}(at)cs.purdue.edu.
[†]Dept. of Computer Science, Univ. of California, Irvine, goodrich(at)acm.org.
[‡]École Polytechnique, stanislas.polu(at)polytechnique.fr.

In fractional cascading [11, 12], we are given a bounded-degree* *catalog graph* $G$, such that each vertex $v$ of $G$ stores a catalog $C(v) \subset U$, for a total order $U$. The catalogs stored at nodes in $G$ are assumed to be stored sorted according to the total order $U$ (and we note that this assumption can be relaxed (for example, for sets of disjoint line segments), so that we assume that the $C(v)$'s contained in any possible search path all belong to a common total order. That is, the partial order defined by all of the $C(v)$'s has a common linearization. Intuitively, a catalog graph $G$ is a data structure and the paths in $G$ are potential traversals in $G$ that might be needed in order to answer a given query. Given a value $x$ belonging to the total order for a path $P$ in $G$, a query for $x$ in $P$ searches for $x$ in the catalog $C(v)$ for each vertex $v$ in $P$. If insertions and deletions are allowed in the $C(v)$'s, then we have the "dynamic fractional cascading" [18] problem. Static fractional cascading solutions due to Chazelle and Guibas [11, 12] allow for queries to be performed in a path of length $k$ in time $O(\log n + k)$, where $n$ is the total size of all the catalogs, and dynamic fractional cascading solutions due to Mehlhorn and Näher [18] show that such queries can be done in a dynamic setting in $O(\log n + k \log \log n)$ time, with updates taking $O(\log n \log \log |U|)$ amortized time. Thus, achieving a dynamic data fractional cascading data structure has an increased complexity. Moreover, this complexity seems inherently difficult to eliminate in the worst case, as it is based on the use of fairly sophisticated data structures that seem necessary in order to handle updates to adjacent catalogs that are very different from one another. The reduced efficiency of dynamic fractional cascading seems to come from its need to dynamically handle discrepancy. Our interest in this paper, therefore, is to address discrepancy head on—to design a scheme for dynamic fractional cascading that is *discrepancy sensitive*. The motivation for such an approach is that there are a number of applications, motivated by real-world scenarios, where the discrepancies between adjacent catalogs are not that great.

## 1.1 Real-World Nearest Neighbor Queries

Suppose, for example, that in a given region, such as a downtown area or university campus, there are sensors that keep track of different physical entities, such as police kiosks, doctors' offices, or coffee shops. When the services of one of these entities are suddenly and urgently needed—e.g., a robbery is in progress, someone is having a heart attack, or a paper deadline is looming—there is a need to quickly compute the position of the entity closest to the location of the sudden emergency event. The important observation we make for such real-world scenarios is that service centers, such as these, are naturally distributed in a fairly uniform way. So if we were to partition our space according to some reasonable data structuring scheme, and apply fractional cascading, we should expect adjacent catalogs to be fairly similar.

---

*We note that a catalog graph of degree $d > 3$ can be transformed into a degree-3 catalog graph by replacing high-degree nodes with complete binary trees.

This paper therefore deals with the questions of how to make such notions of similarity precise and to design structures that store and dynamically update the positions of the entities so that the above-mentioned nearest-neighbor queries can be efficiently processed. The data structures we seek should handle arbitrary insertions and deletions. This is necessary in order to model situations where an entity can suddenly become unavailable or available: for instance, a doctor who is no longer on call or has returned from caring for a patient, a police officer who is busy handling a robbery, or a coffee shop that has run out of espresso. More formally, our motivating application can be stated as follows: Given a dynamic set $S$ of $n$ points in the plane of real coordinates $x$ and $y$, we seek to maintain a data structure that is (i) space-efficient and (ii) supports fast updates (insertions and deletions), as well as fast exact nearest-neighbor (NN) queries. The query points are not necessarily in $S$, but we allow our structure to run faster when $S$ can be partitioned into subsets with low relative local discrepancy.

*Previous Related Work.* There is a considerable amount of prior related work on discrepancy theory and fractional cascading data structures. For prior results in discrepancy theory, for example, please see the excellent book by Chazelle [10]. Subsequent to the introduction of fractional cascading by Chazelle and Guibas [11, 12] and its dynamic implementation by Mehlhorn and Näher [18], there have been many specific uses for this technique, as well as a generalization, due to Sen [23], based on randomized skip lists, and an extension for I/O efficiency due to Yap and Zhu [26].

The prior work on nearest neighbor structures is vast; for more detailed reviews, see the surveys by Alt [1] or Clarkson [13]. Indeed, let us focus here on prior work for planar point sets. For static data, there are several ways to achieve $O(\log n)$ time for nearest-neighbor queries in the plane, including constructing a planar point location data structure "on top" of a Voronoi diagram (e.g., see [22]). For uniformly distributed data, Bentley, Weide, and Yao [6] give optimal algorithms for static data, and Bentley [4] gives an optimal algorithm for the semidynamic (deletion only) case. We are not familiar with any previous optimal fully dynamic algorithms for exact nearest-neighbor queries in uniformly distributed data. For approximate nearest-neighbor queries, Arya *et al.* [3] give an optimal static structure, and Eppstein *et al.* [15] give an optimal dynamic structure. Finally, for general exact nearest-neighbor queries, Chan [8] gives a dynamic method that achieves polylogarithmic expected times for updates and queries. In addition, there has been some work on nearest-neighbors in non-Euclidean settings for "reasonably separated" uniform point sets (e.g., see [7, 17, 16]), but this does work does not apply efficiently to Euclidean metrics on point sets taken from continuous uniform distributions.

*Our Results.* In this paper, we introduce a study of a discrepancy-sensitive approach to dynamic fractional cascading. Unlike the Mehlhorn-Näher approach, which assumes a worst-case distribution for the discrepancies between adjacent catalogs, our approach is sensitive to these differences. That is, it runs faster through low-discrepancy neighbors and slower through high-discrepancy neighbors. We show, for example, that a search for a value $x$ in a collection of catalogs, of size at most $n$, stored in vertices of a path $P$ can be done in time $O(\log n + \sum_{(v,w) \in P} \log \delta_{v,w}(x))$, where $\delta_{v,w}(x)$ is the relative local discrepancy at $x$ of the catalogs stored at the nodes $v$ and $w$ in $G$. Such a discrepancy-sensitive result is useful in a number of real-world scenarios, as we show that there are several practical distributions such that the sum of the relative local discrepancies in the catalogs belonging to a path of length $k$ is $O(k)$ with high probability. For example, we use this approach to provide an efficient data structure for dominated maxima searching in a dynamic set of uniformly distributed points in the plane. This, together with the known fact that the expected number of maxima points in an uniformly distributed set $S$ of $n$ points in $\mathbb{R}^2$ is $O(\log n)$, shows that we can construct a dynamic data structure that can answer queries for nearest neighbors in $S$ using any Minkowski metric, where insertions and deletions run in $O(\log^2 n)$ expected time and queries run in $O(\log n)$ expected time, as well. These expectations assume a uniform distribution, but even with real-life (not uniformly distributed) data we experimentally observe it to hold.

## 2 Discrepancy-Sensitive Dynamic Fractional Cascading

As mentioned above, we are interested in this paper in an approach to dynamic fractional cascading that is based on a local notion of discrepancy in catalog graphs.

*Relative Local Discrepancy in Catalog Graphs.* Weisstein [25] defines a notion for *local discrepancy*, which, for an interval $I$, gives a measure of how much the number of points intersecting $I$ differs from the normalized length of $I$. We are, however, interested in the application to dynamic fractional cascading, which involves comparing adjacent catalogs to each other, not arbitrary intervals to catalogs. Suppose, therefore, that $(v, w)$ is an edge in $G$ and that $C(v)$ and $C(w)$ are the catalogs stored respectively at the vertices $v$ and $w$ in $G$. Let us assume, without loss of generality, that $C(v)$ and $C(w)$ both store sentinel values, "$-\infty$" and "$+\infty$," which are respectively the smallest and the largest elements in the common total order to which all catalog elements belong. For any value $x$, and vertex $v$ in $G$, let $\text{pred}_v(x)$ denote the predecessor of $x$ in $C(v)$, that is, the largest element in $C(v)$ less than or equal to $x$. Likewise, let $\text{succ}_v(x)$ denote the successor of $x$ in $C(v)$, that is, the smallest element in $C(v)$ greater than or equal to $x$. For any edge $(v, w)$ in $G$, we define the *relative local discrepancy* from $C(v)$ to $C(w)$ at $x$ as follows:

$$\delta_{v,w}(x) = |[a,b] \cap C(v)| + |[a,b] \cap C(w)|,$$

where $a = \min\{\text{pred}_v(x), \text{pred}_w(x)\}$ and $b = \max\{\text{succ}_v(x), \text{succ}_w(x)\}$, i.e., the relative local discrepancy from $C(v)$ to $C(w)$ at $x$ is the number of items of $C(v)$ and $C(w)$ falling in the closed interval $[a,b] = [\text{pred}_v(x), \text{succ}_v(x)] \cup [\text{pred}_w(x), \text{succ}_w(x)]$. It is a measure of how different $C(v)$ and $C(w)$ are in the vicinity of $x$. Note that $\delta_{v,w}(x) \geq 2$, even if $C(v) = C(w)$.

*Augmenting a Catalog Graph to Support Searches and Updates.* The main idea of fractional cascading [11, 12] is to augment a catalog graph $G$ with auxiliary structures that support efficient searches and, in the dynamic case [18], updates. The name "fractional cascading" comes from the fact that an effective way to perform this augmentation is to merge fractional samples from the catalogs. Our approach continues this tradition, but implements it in a more localized way.

Let us first give some intuition about our augmentation. Imagine that we have a deterministic skip list [21] built "on top" of the elements in $C(v)$ and that the nodes in this structure are all colored black. Likewise, imagine that we have a deterministic skip list built "on top" of the elements in $C(w)$ and that the nodes in this structure are all colored white. These structures allow for both top-down and bottom-up searches and updates to be performed in $O(\log n)$ time [21]. Now imagine further that we merge these two structures into a common structure by having each black node "cut" any white edge (i.e., interval of white nodes) that it is contained in and having each white node "cut" any black edge that it is contained in. Let us then link the roots of all the remaining bottom-level skip lists. The remaining structure is the "fractionally-cascaded" merge of $C(v)$ and $C(w)$ and this is the structure that we will maintain dynamically.

More formally, our structure is defined so that we maintain the following substructures for each edge $(v, w)$ in $G$ (see Fig. 1):

- We maintain in a "black" deterministic skip list each maximal contiguous interval of $C(v)$ that contains no elements of $C(w)$.

- We maintain in a "white" deterministic skip list each maximal contiguous interval of $C(w)$ that contains no elements of $C(v)$.

- We maintain black-white links between the roots of these skip lists.

- Each bottom-level skip-list interval that is cut by a skip list of the other color has a link to and from the root of that skip list.

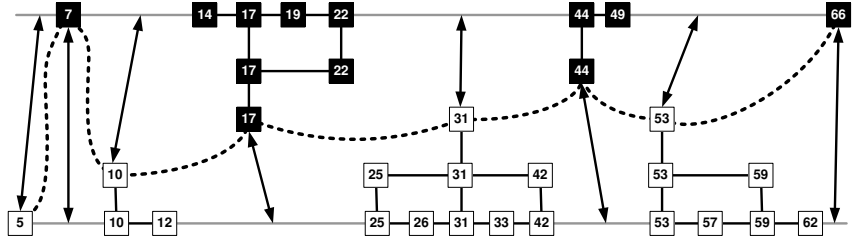

Figure 1: An example of the fractionally-cascaded structures that join a "black" $C(v)$ to a "white" $C(w)$. Skip-list edges are shown in bold, with those cut by a sublist of the opposite colored gray. The links between skip-list roots are shown dashed and the arrowed lines show the links between bottom-level skip-list edges and the roots of the opposite-color skip lists that cut that edge.

*Searches.* A search in a catalog graph $G$ consists of an element $x$ for which we would like to find $\text{pred}_v(x)$ in $C(v)$ for each node $v$ in a given path $P = (v_1, v_2, \ldots, v_k)$. We assume that we have a complete deterministic skip list for the first node, $v_1$, of $P$. This allows us to locate $\text{pred}_{v_1}(x)$ in $O(\log n)$ time, where $n$ is the maximum size of any catalog. For locating $x$ in $C(v_{i+1})$, for $i = 1, \ldots, k-1$, we start from a pointer to $\text{pred}_{v_i}(x)$, which we will have found inductively. There are two cases at this point:

- Case 1: $x$ falls inside a maximal skip list in $C(v_i)$. In this case, we traverse up the skip list for this interval in $C(v_i)$ to its root and then follow the pointer from the root to the interval in $C(v_{i+1})$ containing $x$.

- Case 2: $x$ falls outside a maximal skip list in $C(v_i)$. In this case, we follow the pointer from the "cut" interval in $C(v_i)$ containing $x$ to the root of the skip list in $C(v_{i+1})$ falling in this interval. We then search down this skip list to locate the predecessor of $x$ in $C(v_{i+1})$.

Note that, in either case, each step $i$ of the search, after the first, runs in $O(\log \delta_{v_i, v_{i+1}}(x))$ time, since the size of the skip list we search in for either case is $O(\delta_{v_i, v_{i+1}}(x))$.

*Updates.* Let us consider how to perform an update in our structure, that is, an insertion or deletion in a $C(v)$ list, assuming we have already located the place in $C(v)$ where the update is to occur (let us account separately for the time needed to find this location). We perform the necessary updates for each edge $(v, w)$, of which there are only a constant number, according to the following cases:

- **Insert** $y$:
    - Case 1: $y$ falls inside a maximal skip list $L$ in $C(v)$. In this case, we simply insert $y$ in $L$.
    - Case 2: $y$ falls outside a maximal skip list in $C(v)$. In this case, we follow the interval pointer from the (gray) interval in $C(v)$ containing $y$ to the skip list $L$ in $C(w)$ and search down for $y$ in this list. If $y$ falls in the interior of $L$ then we split $L$ at $y$, set up $y$ as its own skip list in $C(v)$ and update the pointers of the three new root nodes. If $y$ falls outside $L$, then we simply insert $y$ in the appropriate predecessor or successor skip list in $C(v)$ and update the (gray) interval to now have $y$ as an endpoint.

- **Delete** $y$:
    - Case 1: $y$ falls in a maximal skip list $L$ in $C(v)$ with at least one other element. In this case, we simply remove $y$ from $L$ (possibly updating boundary pointers if $y$ was the smallest or largest element in $L$ or the root pointers, if $y$ was a root element— so that the appropriate adjacent pointers now point to the new root of $L$).
    - Case 2: $y$ is the only element of its skip list in $C(v)$. In this case, we follow the pointers from $y$'s (root) node to the two skip lists

in $C(w)$ that $y$ separates, and we perform a splice of these two structures, updating the root pointers as needed.

Note that in either an insertion or a deletion, the time needed to perform all the necessary local searching, insertions, deletions, splits, and/or splices is $O(\log \delta_{v,w}(y))$.

**Theorem 1** *A catalog graph $G$, with maximum catalog size $n$, can be augmented with additional structures so as to support searches for an element $x$ in the catalogs in a path $P$ in $G$ in time $O(\log n + \sum_{(v,w) \in P} \log \delta_{v,w}(x))$. Likewise, a sequence of updates for an element $y$ in catalogs in a path $P$ in $G$ can be done in these structures in time $O(\log n + \sum_{(v,w) \in P} \log \delta_{v,w}(y))$.*

*Additional Analysis.* So as to better motivate the use of relative local discrepancy as a performance parameter, we provide in this subsection some additional analysis of our dynamic fractional cascading solution.

*Uniform data.* Suppose that each catalog in $G$ contains $n$ points chosen independently and uniformly at random from the interval $[0, 1]$. In this case, the set of points in a catalog $C(v)$ define a set of order statistics, and the distribution of the length of consecutive spacings therefore follows the Beta distribution with parameters 1 and $n$ (e.g., see [2, 14]). Thus, the expected interval length is $1/(n+1)$. Having fixed such an interval in $C(v)$, the number of points in $C(w)$ that falls in this interval follows a Binomial distribution, with probability equal to the length of the interval. Thus, the distribution of each $\delta_{v,w}(v)$ follows the Beta-Binomial distribution, with parameters 1 and $n$, which has expected value $\mu = n/(n+1)$ [24].

The performance of searching and updating our augmented structures at an element $x$ along a path $P = (v_1, \ldots, v_k)$ in a catalog graph $G$ depends on the random variable,

$$T_P = \sum_{(v_i, v_{i+1}) \in P} \log \delta_{v_i, v_{i+1}}(x).$$

Unfortunately, the relative local discrepancies for consecutive edges in $P$ are not necessarily independent. Even so, we can write

$$T_P = \sum_{(v_i, v_{i+1}) \in P,\ \text{odd } i} \log \delta_{v_i, v_{i+1}}(x) + \sum_{(v_i, v_{i+1}) \in P,\ \text{even } i} \log \delta_{v_i, v_{i+1}}(x), \quad (1)$$

and we note that each term in the separate sums are independent. Thus, we can bound the degree to which $T_P$ differs from its expectation by adding bounds on the two sums. Combining this with the expected value of the associated Beta-Binomial distribution given above, we can use a Chernoff bound twice (e.g., see [20]) to prove the following (we give the proof in the final version):

**Theorem 2** *Given a catalog graph $G$ such that each catalog is a set of $O(n)$ independent, uniform random points in the interval $[0, 1]$, then for any path $P$ of length $k$ in $G$, $\sum_{(v,w)\in P} \log \delta_{v,w}(x)$ is $O(k)$ with probability $1 - 1/2^k$.*

Using this result, we can take the dynamic range searching structure of Mehlhorn and Näher [18], which is based on range trees (e.g., see [22], and replace their dynamic fractional cascading solution with ours, which gives us the following:

**Theorem 3** *We can maintain a dynamic range searching data structure for a set of points taken uniformly at random in the unit cube so as to support point insertions and deletions in $O(\log n)$ time w.h.p. and the reporting of all the points in a rectangular query range $[x_1, x_2] \times [y_1, y_2]$ in $O(\log n + k)$ time w.h.p., where $k$ is the number of points returned by the query.*

Our data structure is deterministic, and works in the standard comparison-based pointer-machine model. It therefore matches w.h.p. the range-searching query and update times of Mortensen [19], which are instead for the RAM model and make use of bit twiddling. The high-probability bound in Theorem 3 comes from Theorem 2.

*Other Upper bounds.* Given the above proof technique derived for the uniform data case, we can motivate bounds on $T_P$ for other distributions by probabilistically bounding of each of the two terms in Equation 1. That is, let us concentrate on the odd $i$ case (the analysis for even $i$ being similar) and let us consider a random variable $X = X_1 + \cdots + X_k$ such that, for each $i = 1, \ldots, k/2$, $\Pr(\log \delta_{v_{2i-1}, v_{2i}}(x) \leq y) \leq \Pr(X_i \leq y)$. Then we can bound the odd summation with a bound for $X$. In particular, we can use various Chernoff bounds to show that $X$ is $O(k)$ with high probability for each of the following cases:

- Each $X_i$ is Binomial with constant expected value.

- Each $X_i$ is geometric (the discrete counterpart to the exponential distribution).

- Each $X_i$ is Poisson with bounded expected value.

Having provided our general framework for discrepancy-sensitive dynamic fractional cascading, let us give a concrete application to nearest-neighbor searching.

## 3 Dynamic Dominated Maxima

This section describes a scheme for dynamically maintaining a set $S$ of points drawn from a uniform distribution in a rectangle, so that a *dominated maxima* query can be done in $O(\log n)$ expected time: Given a query point $q$, the query returns the set of maximal elements among the points of $S$ that are dominated by $q$; note that the expected size of the output is itself $O(\log n)$ (because of the uniform distribution). The expected time for an update will be shown to be $O(\log^2 n)$.

We shall find it necessary to maintain 4 such data structures, one for each of the 4 possible sets of coordinate axes obtained by reversing the direction of {neither,one,both} of the $x$ and $y$ axes – having all 4 such structures makes it possible to achieve the bounds we claim but imposes only a constant factor of 4 on the complexity bounds.

In order to more explicitly define the 4 above-mentioned problems, and also to facilitate the understanding of our algorithm, we will consider the smallest origin-centered square containing the whole set $S$ for a given state of $S$. We position four coordinate systems, one at each of the four corners of the square, with the origin being at the corresponding corner and the directions of the axes pointing from the origin along the edges of the square. We call these four coordinate systems *South-West* (abbreviated as $SW$), *South-East* ($SE$), *North-West* ($NW$), *North-East* ($NE$). For a point $q \in S$, we use $x_{SW}(q)$ (resp., $y_{SW}(q)$) to denote the $x$ (resp., $y$) coordinate of $q$ in the $SW$ coordinate system. A similar notation is used for the other three coordinate systems. Such coordinate systems and point coordinates are depicted in Fig. ??.

The 4 problems mentioned above are then the following: (i) A South-West problem that pertains to the subset of $S$ that is dominated by the query point $q_0$ in the $SW$ coordinate system, i.e., the subset "below and to the left of $q_0$"; (ii) a South-East problem that pertains to the subset of $S$ that is dominated by the query point $q_0$ in the $SE$ coordinate system (the subset "below and to the right of $q_0$"); (iii) a North-East problem that pertains to the subset of $S$ that is dominated by the query point $q_0$ in the $NE$ coordinate system (the subset "above and to the right of $q_0$"); and (iv) a North-West problem that pertains to the subset of $S$ that is dominated by the query point $q_0$ in the $NW$ coordinate system (the subset "above and to the left of $q_0$").

Recall that a point $q$ is *maximal* in the set $S$ relative to the $SW$ coordinate system iff for every other point $q' \in S$ at least one of the following

inequalities holds:

$$x_{SW}(q') \leq x_{SW}(q) \qquad y_{SW}(q') \leq y_{SW}(q),$$

which, in words, can be stated as: "no other point of $S$ dominates $q$ in the $SW$ coordinate system." For a point $q$ and a set $S$ we also define the notion of a maximal set in the $SW$ coordinate system with respect to $q$. This set, denoted by $M_{SW}(S,q)$, is computed by first considering only those points in $S$ that are dominated by $q$ in the $SW$ coordinate system (i.e., the subset of $S$ below and to the left of $q$) and then computing the maximal points of that subset. All points in $M_{SW}(S,q)$ are assumed to be sorted by increasing $x$ coordinates. A similar notation is used for the other three coordinate systems.

In the rest of our discussion we focus on the South-West problem. All of our solutions for this South-West problem can be translated into similar ones for the South-East, North-East, and North-West problems.

## 3.1 The Data Structure

Let $T_x$ be an $n$-node search tree structure whose nodes are the $n$ points of $S$ ordered by their $x$ coordinates. $T_x$ verifies the following properties, $v$ being a node of $T_x$ :

- $T_x$ is a weight balanced binary search tree
- All nodes in the right subtree of $v$ have greater $x$ value than $v$
- All nodes in the left subtree of $v$ have lesser value than $v$

For each node $v$ in $T_x$, we use $Sl_v$ to denote the subset of $S$ that lies in the subtree of $v$ and have $x$ coordinate lesser or equal to $v$'s one. Each such $Sl_v$ is itself organized as a dynamic search structure according to the $y$ coordinates of the points in it. The $T_x$ tree and its associated $Sl_v$'s are organized as the dynamic fractional cascading structure described above. With this structure in place, for every path $\mathcal{P}$ in $T_x$, searching for $y_0$ in $Sl_v$ for every $v \in \mathcal{P}$ can be done in $O(\log n + |\mathcal{P}|)$ expected time.

An update to this structure due to insertion or deletion of a point consists of adding or removing a node of $T_x$, updating all the $Sl_v$ sets from that node to the root and finally then rebalancing $T_x$. Note that the insertion of a point $(x_0, y_0)$ does not cause the creation of a new node in $T_x$ if there exists already a point with $x_0$ coordinates, but only an update in the underlying dynamic fractional cascading structure. We have the equivalent property for deletion. Rebalancing the tree implies $O(1)$ rotations. A rotation associated with three node $v, v', v''$ implies the reconstruction of the underlying sets $Sl_v, Sl'_v, Sl''_v$, that is, $O(|Sl_v|)$ insertions and deletions in the dynamic fractional cascading structure. Since $T_x$ is a weight balanced search tree, the amortized value of $|Sl_v|$ is $\log n$. Thus an update to this structure takes $O(\log n)$ amortized time.

In addition to the above, each copy of a point $q$ in $Sl_v$ stores the following:
- $l_{SW}(v, q)$ = the leftmost (hence, highest) point in $M_{SW}(S_v, q)$.
- $r_{SW}(v, q)$ = the rightmost (hence, lowest) point in $M_{SW}(S_v, q)$.
- $l_{SE}(v, q)$ = the leftmost (hence, lowest) point in $M_{SE}(S_v, q)$.
- $r_{SE}(v, q)$ = the rightmost (hence, highest) point in $M_{SE}(S_v, q)$.
- $l_{NW}(v, q)$ = the leftmost (hence, lowest) point in $M_{NW}(S_v, q)$.
- $r_{NW}(v, q)$ = the rightmost (hence, highest) point in $M_{NW}(S_v, q)$.
- $l_{NE}(v, q)$ = the leftmost (hence, highest) point in $M_{NE}(S_v, q)$.
- $r_{NE}(v, q)$ = the rightmost (hence, lowest) point in $M_{NE}(S_v, q)$.

The above quantities will be shown to facilitate a query, but they also impose the burden of dynamically updating them. We need to describe how a query is processed, and how to dynamically update all of the above quantities.

### 3.2 Processing a Query

The query processing consists of, given a query point $q_0$, returning the maximal elements of the subset of $S$ dominated by $q_0$ in the $SW$ coordinate system. (The query point is arbitrary and need not be in $S$.)

More formally, to process a query for a point $q_0$ with the coordinates $(x_0, y_0)$, we do the following:

1. First we locate the node which has greatest $x$ value lesser or equal to $x_0$ in $T_x$, thereby defining a root-to-leaf path $\mathcal{P}$ in $T_x$. Let $v_1, \ldots, v_t$ be (in left to right order) the nodes whose right sibling is on $\mathcal{P}$. We henceforth refer to these nodes as the *fringe of $x_0$ in $T_x$*. Note that $t \leq \log n$, and that every point in $\bigcup_{i=1}^{t} Sl_{v_i}$ has an $x$ coordinate that is $\leq x_0$ and that there is no other such points.

2. Within every $Sl_{v_i}$, $1 \leq i \leq t$, let $y'_i$ be the largest $y$ coordinate that is $\leq y_0$. Computing all the $y'_i$s involves locating $y_0$ in every $Sl_{v_i}$. Using the dynamic fractional cascading search structure, the computation of all the $y'_i$s can be done in $O(\log n + t)$ expected time, which is $O(\log n)$.

3. Let $Y_1, \ldots, Y_t$ be defined inductively as follows:

    (a) $Y_t = -\infty$

(b) $Y_{k-1} = \max\{Y_k, y'_k\}$ for $k = t-1, t-2, \ldots, 1$.

In words, $Y_k$ ($k < t$) is the largest $y$ coordinate among the points in $\bigcup_{i=k+1}^{t} Sl_{v_k}$.

4. Enumerate the points in $M_{SW}(S, q)$. Before explaining how this enumeration done, we point out that the point of $S$ that constitutes the South-West solution must belong to $M_{SW}(S, q)$, which is easy to prove by contradiction. We also point out that the expected number of points in $M_{SW}(S, q)$ is $O(\log |S|)$, hence $O(\log n)$. Thus, the $O(\log n)$ average query performance would be achieved if we could somehow enumerate the points of $U = M_{SW}(S, q)$ in time $O(|U|)$. We do this by first observing that the subset of $S$ from which the maximal points are computed consists of the subset of $\bigcup_{i=1}^{t} Sl_{v_k}$ having $y$ coordinates $< y_0$. Our strategy will be to enumerate, in the order $k = 1, \ldots, t$ the maximal points of $Sl_{v_k}$ that belong to $U$, call their set $U_k$, *stopping as soon as the about-to-be-enumerated $y$ coordinate drops below $Y_k$*. (If we did not stop at that point, we would be enumerating points that do not belong to $U$.) This enumeration of $U_k$ is done as follows:

   (a) Let $q_k$ be the point with the $y$ coordinate $y'_k$ (that is, $q_k$ is the highest point of $Sl_{v_k}$ whose $y$ coordinate is $\leq y_0$).

   (b) While the $y$ coordinate of $q_k$ is $\geq Y_k$, we (i) include $q_k$ as a member of $U_k$, and then (ii) set $q_k = r_{SE}(v, q_k)$, which is the rightmost (hence, highest) point in $M_{SE}(S_v, q_k)$.

   Of course, in the above, $U$ is the concatenation of $U_1, \ldots, U_t$.

5. Since we have not checked the points with $y$-coordinate equal to $y_0$ in $M_{SW}(S, q)$, we need to add them to $U$. This can be done by searching for $y_0$ in the fringe of $x_0$ which takes $O(\log n)$ expected time using the fractional cascading structure.

As argued above, the average complexity of the above query processing is $O(\log n)$. We now turn our attention to the dynamic updates. We begin with the case of insertions.

## 3.3 Processing an Insertion

Let $q_0 = (x_0, y_0)$ be the point being inserted. We already argued that the fractional cascading structure can be updated in $O(\log n)$ expected time as a result of this insertion. The main task we face now is how to update the quantities $l_{SW}(v,q)$, $r_{SW}(v,q)$, $l_{SE}(v,q)$, $r_{SE}(v,q)$, $l_{NW}(v,q)$, $r_{NW}(v,q)$, $l_{NE}(v,q)$, and $r_{NE}(v,q)$, for each $q = (x,y) \in S$ and each $v$ that is ancestor of $x$ in $T_x$. We explain how to update only $r_{SE}(v,q)$ for all $v$'s that are ancestors of $x$ in $T_x$; similar updating can be repeated for each of the seven other quantities (relative to their own frame of reference).

We begin with the updating of the $r_{SE}(v,q)$'s for all points other than $q_0$ (i.e., the points in $S - \{q_0\}$). And we will explain how to compute the $r_{SE}(v, q_0)$ separately.

The first step is to compute, as a query that is processed just as in the previous section (except that the coordinate system is different), the set $U = M_{NE}(S, q_0)$, where, as before, the expected size of $U$ is $O(\log n)$. The only points $q$ of $S$ whose $r_{SE}(v,q)$ may change are in $U$. For each point $q$ of $U$, we update its (at most $\log n$) $r_{SE}(v,q)$ values. This is done in constant time for each value, by checking whether $q_0$ can cause an improvement when $v$ is ancestor of $q_0$. The total update time for doing this is therefore $O(|U| \log n)$, which is $O(\log^2 n)$ on average.

To compute the $r_{SE}(v, q_0)$, we first compute $U' = M_{SE}(S, q_0)$ as a query, hence in $O(\log n)$ expected time. We then walk along the path from $x_0$ to the root in $T_x$, and at each node $v$ along this path we set $r_{SE}(v, q_0)$ equal to the highest point of $U'$ that is in $Sl_v$. Note that this whole walk can be done in time $O(\log n)$ because of monotonicity: The $Sl_v$'s of the nodes on that walk to the root monotonically "swallow" $U'$ in left-to-right order (hence, by increasing $y$ coordinates). Thus we end up going through $U'$ only once (not $\log n$ times).

### 3.4 Processing a Deletion

Let $q_0 = (x_0, y_0)$ be the point being deleted. We already argued that the fractional cascading structure can be updated in $O(\log n)$ expected time as a result of this deletion. Now we need to show how to update the quantities $l_{SW}(v,q)$, $r_{SW}(v,q)$, $l_{SE}(v,q)$, $r_{SE}(v,q)$, $l_{NW}(v,q)$, $r_{NW}(v,q)$, $l_{NE}(v,q)$, and $r_{NE}(v,q)$, for each $q = (x,y) \in S$ and each $v$ that is ancestor of $x_0$ in $T_x$. We explain how to do it for $r_{SE}(v,q)$ for all $v$ that are ancestors of $x_0$ in $T_x$, all other values are updated similarly (relative to their own frame of reference).

First, we compute each of the sets $U = M_{NW}(S, q_0)$ and $U' = M_{SW}(S, q_0)$ as queries (and, hence, in $O(\log n)$ expected time). The only points $q$ of $S$ whose $r_{SE}(v,q)$ may change as a result of the deletion are in $U$. Moreover, for each such point $q$ whose $r_{SE}(v,q)$ changes, its new $r_{SE}(v,q)$ is either in $U'$ or it is the old $r_{SE}(v, q_0)$. The best candidate from $U'$ for each $q \in U$ need not be done in isolation; rather, it can be done for all the points of $U$ together. This can be performed in a manner reminiscent of the way two sorted lists are merged, by walking simultaneously along $U$ and $U'$. This has to be done only once (not repeated for the $r_{SE}(v, q)$ of every ancestor $v$ of $x_0$). On the other hand, the comparison of the old $r_{SE}(v, q)$ with the two new candidates, which are the old $r_{SE}(v, q_0)$, and the point of $U'$ determined during the above-mentioned merge-like procedure, needs to be done for every $v$ and $q$. Hence, the overall time for a deletion is $O(\log^2 n)$ on average.

## 4 Dynamic Nearest Neighbors in Minkowski Metrics

Given a nearest-neighbor query for a point $q_0$, in a set $S$ of uniformly-distributed points in an axis-aligned rectangle, we partition the problem into four sub-problems: (i) a South-West problem that consists of computing the nearest neighbor from among the subset of $S$ that is dominated by the query point $q_0$ in the $SW$ coordinate system, i.e., the subset "below and to the left of $q_0$"; (ii) a South-East problem that consists of computing the nearest neighbor from among the subset of $S$ that is dominated by the query point $q_0$ in the $SE$ coordinate system (the subset "below and to the right of $q_0$"); (iii) a North-East problem that consists of computing the nearest neighbor from among the subset of $S$ that is dominated by the query point $q_0$ in the $NE$ coordinate system (the subset "above and to the right of $q_0$"); and (iv) a North-West problem that consists of computing the nearest neighbor from among the subset of $S$ that is dominated by the query point $q_0$ in the $NW$ coordinate system (the subset "above and to the left of $q_0$"). We solve all of (i)–(iv) and choose, as the solution to the nearest-neighbor query, the best from the four answers they return. Our performance bounds for this problem therefore immediately follow from those we established in the previous section for the dynamic dominated maxima problem: $O(\log n)$ expected query time, and $O(\log^2 n)$ expected time for an update (insertion or deletion).

## 5 Experimental Results

In this section we confront our results holding for evenly distributed sets to real data sets. First we evaluate the local discrepancy distribution along a path down a search tree $T_x$ as we use it in our nearest-neighbor data structure in the case of real and evenly distributed data. Then we show the validity of the expected $log(n)$ size of $M_{SW}(S, q)$ in the case of real data.

*Real Data Set.*

To run our experimentations, we have chosen to use real data extracted from Tiger[†] by Dr. Yufei Tao[‡]. Originally, the data represented areas in Long Beach County (areas that could be seen as RFID readers range). We have kept the center of those areas and stored their $x$ and $y$ coordinates. This data set is interesting because of his good representation of density variations in the different areas of the city, characteristic of human activity. Let denote this real data set of cardinal 53145 by $S$. To avoid under-estimated results induced by border effects (querying from points outside the set or on its border), we have run the tests on points belonging to the real data set $S$ itself and which are located in a restricted inner-area of S of high density.

*Local Discrepancy on a Range Tree.* In this section we explore the distributions of the local discrepancy in the catalogs of the nodes of a range tree, augmented using our dynamic fractional cascading structure.

*Protocol.* To evaluate the distributions of the local discrepancy along a path in the range tree we use, we have inserted the points of the real data set $S$ in such a range tree and chose random query points $(x, y)$. For each point, we calculated the local discrepancy relative to $y$ for each edge on the path from the leaf associated with $x$ to the root of the tree. We also did the same work with the same number of evenly distributed points (see Fig. 2).

*Results.*

As we see in Fig. 2, the distributions of local discrepancy for the real data set is very close to the distributions of local discrepancy in the case of evenly distributed points. Their plot in logarithmic scale indicates that they are very close to exponential distributions, which shows that the demonstration for theorem 3 still holds in the case of the real data set.

*Maxima Points Chains Length.*

In this section we show the validity of the expected $log(n)$ size of $M_{SW}(S, q)$ in the case of real data.

*Protocol.* Our results present an evaluation of the cardinal of $M_{SW}(S, q)$ on the real data set $S$ described above. To be able to evaluate $|M_{SW}(S, q)|$ as a function of $|S|$, we ran the tests on $S$ itself and then on $S$ dominated by random deletions of points. These random deletions preserved the original distribution of $S$ while decreasing its cardinal. Let's denote those dominated sets by $S_k$. The tests were run by querying randomly chosen points $q \in S_k$ for each dominated subsets $S_k$.

---

[†]U.S. Census Bureau - Topologically Integrated Geographic Encoding and Referencing system

[‡]http://www.cs.cityu.edu.hk/ taoyf/ds.html

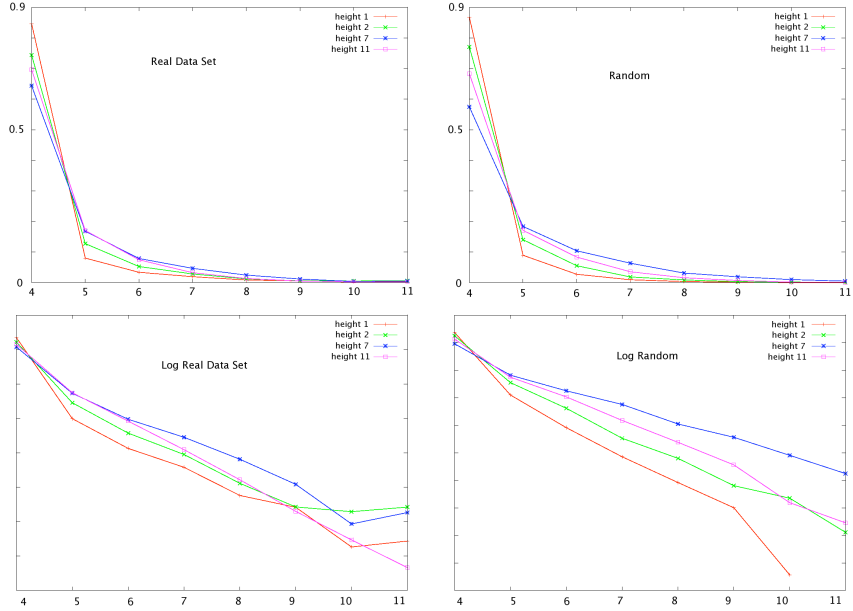

Figure 2: Distributions of the local discrepancy along top-down path in a range tree using real data set in the upper-left corner and evenly distributed points on the upper-right corner. The distribution $height_k$ represents the distribution of local discrepancy for edges between nodes at height $k-1$ and $k$ containing respectively $2^{k-1}$ and $2^k$ points in their catalogs. The two plots below show the same distributions on a log scale.

We also ran the tests in the same conditions on evenly distributed sets $R_k$ of cardinal equal to $|S_k|$.

*Results.* The cardinal of $M_{SW}(S,q)$ is expected to be a $O(log(|S|))$, and more precisely bounded by $h(|S|) = 1 + \sum_{k=1}^{|S|} \frac{1}{k}$ [5] if the points of $S$ are evenly distributed. Fig. 3 shows our experimental results for the sets $S_k$ and $R_k$ compared to $h(|S_k|)$ and $log(|S_k|)$. We see that the assertion holds in the case of the real data points we used, which we think representative of the type of distribution our system could deal with.

The two experimental results put together indicate that the complexity bounds announced for our dynamic nearest neighbor solution still hold in the case of real data set.

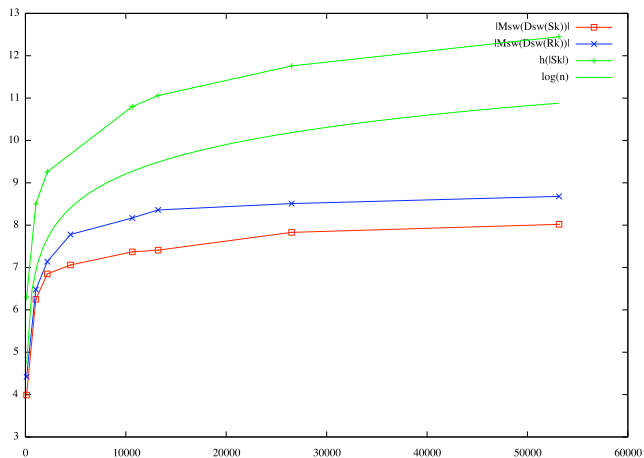

Figure 3: Results for $|M_{SW}(S_k)|$. The x-axis represents the number of point in $S_k$ while the y-axis represents the cardinal of $M_{SW}(S_k)$.

### Acknowledgments

We would like to thank Yaming Yu for pointing us to the Beta-Binomial distribution as an analysis tool for relative local discrepancy of catalogs of uniformly distributed points.

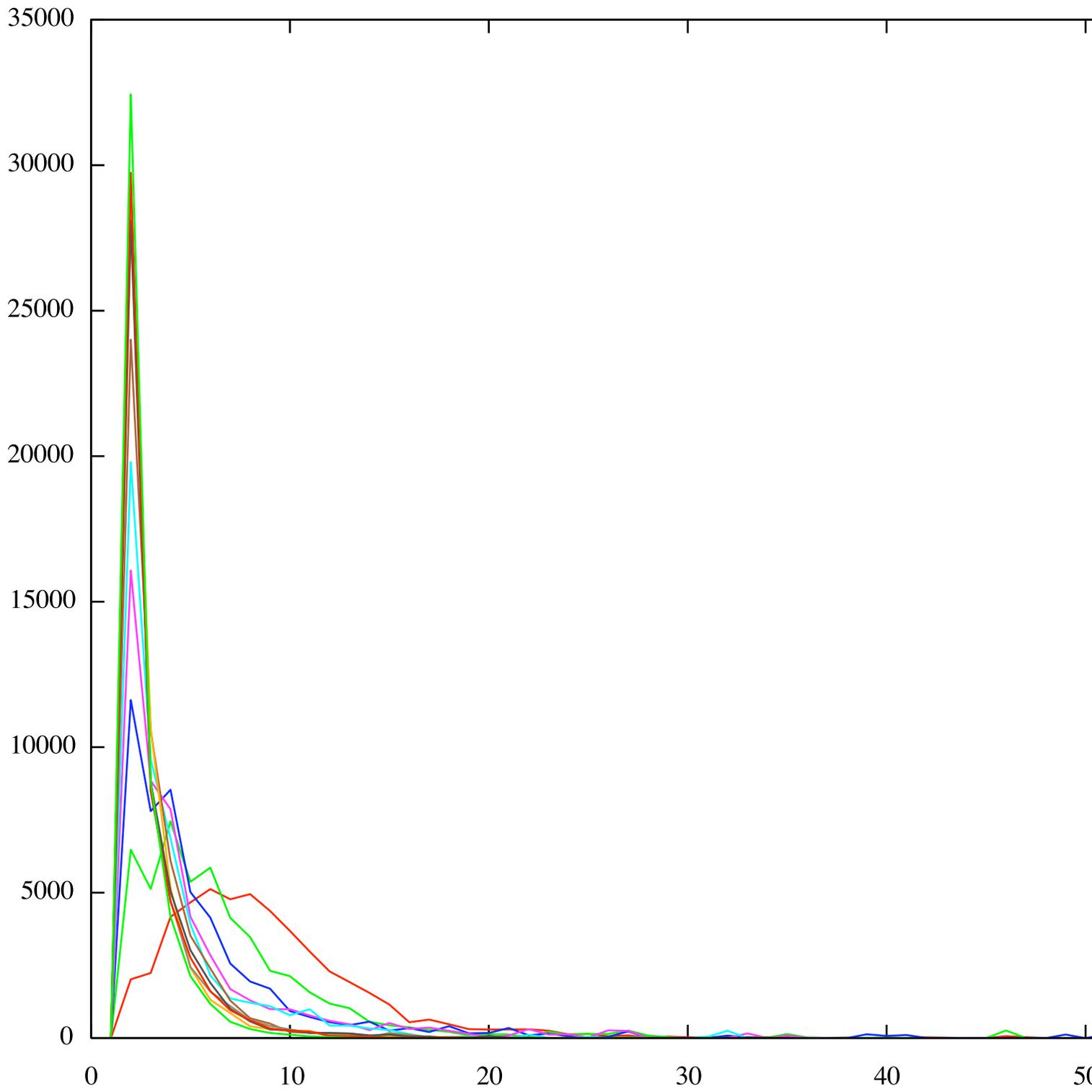